# Entropy jump at the first-order vortex phase transition in $Bi_2Sr_2CaCu_2O_{8+\delta}$ with columnar defects

Rumi G[a], Albornoz L J[a], Pedrazzini P[a], Dolz M I[b], Pastoriza H[a], van der Beek C J[c], Konczykowski M[c], Fasano Y[a1]

[a]*Low Temperature Lab and Instituto Balseiro, Centro Atómico Bariloche, Bustillo 9500, Bariloche (8400), Argentina*
[b]*Physics Department, University of San Luis, Ejército de los Andes 950, San Luis (5700), Argentina*
[c]*Laboratoire des Solides Irradiés, École Polytechnique, Route de Saclay 91128, Palaiseau (91120), France*

**Abstract**

We study the entropy jump associated with the first-order vortex melting transition (FOT) in $Bi_2Sr_2CaCu_2O_{8+\delta}$ crystals by means of Hall probe magnetometry. The samples present a diluted distribution of columnar defects (CD) introduced by irradiation with Xe ions. The FOT is detected in ac transmittivity measurements as a paramagnetic peak, the height of which is proportional to the enthalpy difference entailed by the transition. By applying the Clausius-Clapeyron relation, we quantify the evolution of the entropy jump Δs as a function of the FOT temperature, $T_{FOT}$, in both pristine crystals and crystals with CD. On increasing the density of CD, Δs decreases monotonically with respect to values found in pristine samples. The Δs versus $T_{FOT}$ dependence in the case of pristine samples follows reasonably well the theoretical prediction of dominant electromagnetic coupling for a model neglecting the effect of disorder. The data for samples with a diluted distribution of CD are not properly described by such a theoretical model.

## 1. Introduction

In high-temperature cuprate superconductors, vortex matter exhibits a first-order transition (FOT) [1,2] on cooling from a liquid phase with zero shear modulus [3] to a solid glassy phase. This so-called melting transition is the result of thermal fluctuations overcoming vortex-vortex and vortex-disorder interactions. [1] In the case of layered cuprates such as $Bi_2Sr_2CaCu_2O_{8+\delta}$, flux lines should be considered as stacks of two-dimensional (2D) pancake vortices located in the superconducting $CuO_2$ planes. Pancake vortices belonging to the same vortex line and situated in adjacent $CuO_2$ planes are coupled by electromagnetic and Josephson interactions. [4] Some experimental works [5,6] report on the disordered nature of the vortex solid phase, suggesting that the in-plane structural order of vortex matter might not be the only symmetry broken at the FOT. In addition, Josephson plasma resonance experiments reveal that very close to the FOT the relative displacement of pancake vortices in adjacent $CuO_2$ planes becomes of the order of the lattice spacing. [7] This suggests that the FOT quite likely corresponds to a layer decoupling mediated by the diffusion onset of pancake vortices. The order parameter suddenly changing at the FOT transition would be the mean gauge-invariant phase difference between layers rather than the transversal order of the flux line structure. This is supported by the absence of melting point depression in vortex nanocrystals of $Bi_2Sr_2CaCu_2O_{8+\delta}$ with less than hundred flux lines but tens of thousands of pancake vortices. [8]

Phenomenologically, the quantity abruptly changing at the FOT is the vortex density: the magnetic induction $B$ in the solid phase jumps to a larger value when passing to the liquid phase. The enthalpy change at the FOT is proportional to this jump, $\Delta B$, and considering the Clausius-Clapeyron relation we obtain the modulus of the entropy jump per pancake vortex [9]

$$\Delta s = \frac{\Phi_0 d}{4\pi} \frac{|\Delta B|}{B_{FOT}} \left| \frac{dH_{FOT}}{dT_{FOT}} \right| \qquad (1)$$

where $d$ is the spacing between adjacent $CuO_2$ planes, $\Phi_0$ the magnetic flux quantum, and $H_{FOT}$, $T_{FOT}$, and $B_{FOT}$ the applied field, temperature and magnetic induction at which the FOT takes place. In the case of

---

vortex matter in layered superconductors, magnetization experiments suggest that the entropy jump diverges on approaching $T_c$.[2,9] These observations were explained by a theoretical work considering the electromagnetic and Josephson coupling between pancake vortices of adjacent superconducting planes, the temperature dependence of the Ginzburg-Landau coherence length and penetration depth, $\lambda$, and the shape of the FOT line for a given material [10]. This theory takes into account the crossover from an electromagnetic- towards a Josephson-dominated regime induced near $T_c$ on reaching the condition $d\gamma < \lambda(T)$, where $\gamma$ is the anisotropy. Even if this theory does not take into account the effect of disorder, some of us showed that in the case of several optimally-doped pristine $Bi_2Sr_2CaCu_2O_8$ samples $\Delta s$ diverges close to $T_c$ following the temperature-evolution predicted for dominant electromagnetic coupling [9].

Here we unveil the effect of disorder on the temperature-evolution of the entropy-jump at the FOT by introducing a diluted distribution of strong, correlated pinning centers in the form of columnar defects (CD) induced by irradiation with high-energy Xe heavy ions. We study the case of two CD densities $n_d$, sufficiently low so as for the FOT transition to persists to fields of up to 60 Oe. We find that there is a gradual decrease of the entropy jump on increasing $n_d$. We also show that the temperature-evolution of $\Delta s$ for a given density of CD falls apart from the functionality found for pristine samples in a way that cannot be explained by an enhancement of the relevance of Josephson coupling.

## 2. Experimental

We study $Bi_2Sr_2CaCu_2O_{8+\delta}$ single crystals with a random, dilute distribution of CD's generated by irradiation with 1 GeV Xe ions performed at GANIL, France. We measure samples with matching fields $B_\Phi \equiv n_d\Phi_0$=10 and 30G; note that for $B=B_\Phi$ the number of vortices equals that of CD's. Data in four pristine optimally-doped $Bi_2Sr_2CaCu_2O_{8+\delta}$ samples [9] are also presented for comparison. We apply local Hall-probe magnetometry to characterize the FOT in these samples.

We measure the sample stray field $B$ using 16 x 16μm$^2$ active area two-dimensional electron-gas sensors (see inset to Fig. 1). We detect and characterize the FOT by means of an *ac* technique. A static magnetic field $H$ plus a ripple field $h_{ac}$ are applied along the *c*-axis upon cooling from the normal state (field cooling). [9] By means of a DSP lock-in technique we record both magnitude and phase of the Fourier component of $B$ at the frequency of the $h_{ac}$ excitation. The in-phase component of the signal as a function of temperature is proportional to the local susceptibility and is therefore a suitable magnitude to detect abrupt changes in $B$. We normalize this in-phase component setting its low- and high-temperature limiting values to zero and one, respectively, in order to obtain the sample transmittivity *T'*.[9] In our measurements the ripple field peak-amplitude lies in the range 0.7-1.5 Oe and the frequency is set between 3.1 to 7.1 Hz.

## 3. Results and discussion

Figure 1 shows transmittivity data *T'* as a function of temperature at different applied fields for the case of heavy-ion irradiated $Bi_2Sr_2CaCu_2O_8$ with a $B_\Phi$ = 30 G. The transmittivity is constant and equal to 1 in the normal state (high temperatures); on cooling into the superconducting phase, a paramagnetic peak develops, and at low temperatures superconducting screening dominates the signal. On increasing $H$ the peak moves to lower temperatures, confirming the melting transition of vortex matter even in the presence of strong disorder. The temperature at which the peak is detected is robust with respect to changes in the amplitude and the frequency of $h_{ac}$, indicating the transition is truly first-order.[9] We considered the transition temperature $T_{FOT}$ as that corresponding to the maximum of the paramagnetic peak, see Fig. 1 (a). The *T'* curves show paramagnetic peaks up to roughly 60 Oe; for larger fields the peaks are washed out by the sample screening. A similar phenomenology is observed in the $B_\Phi$=10 G sample, indicating that in $Bi_2Sr_2CaCu_2O_8$ with a dilute distribution of CD's the FOT persists up to at least 60 Oe within our experimental resolution.

Figure 1 (a) indicates the height of the paramagnetic peaks, *ΔT'*, proportional to *ΔB* at the FOT according to [11]

$$\Delta T' = \frac{2\,\Delta B}{\pi h_{ac}} \quad\quad\quad\quad\quad\quad\quad\quad\quad\quad\quad\quad\quad\quad\quad\quad\quad (2)$$

The entropy jump $\Delta s$ can therefore be obtained by extracting the value of $\Delta B$ from the height of the peaks following Eq. 2, and estimating the derivative of $H_{FOT}$ with respect to the $T_{FOT}$ line obtained from the location of the paramagnetic peaks, see Eq. 1. The entropy jumps for samples with $B_\Phi$=10 and 30 G obtained in this way are shown in Fig. 1 (b) with full symbols. The figure also shows results for four optimally-doped pristine samples that fall in a single bunch of data (open symbols). The divergence of $\Delta s$ on approaching the critical temperature is observed in pristine as well as in samples with CD. However, at low temperatures, $\Delta s$ for samples with disorder decreases on cooling at a faster rate than for pristine samples. The rate of decrease is enhanced as the density of CD and therefore the level of disorder of the vortex system increases.

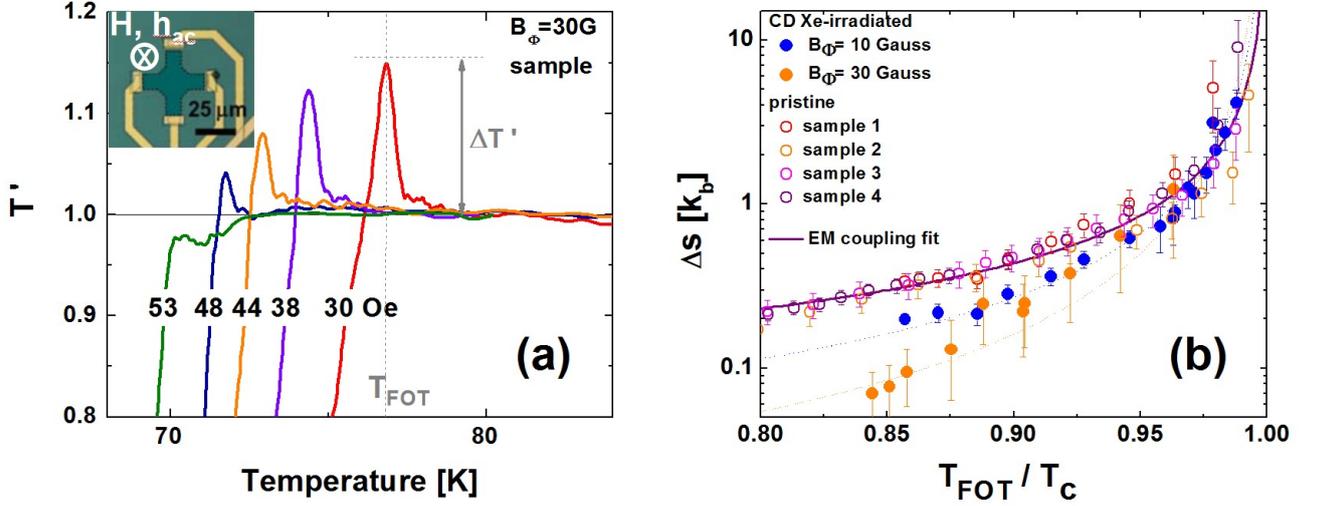

Figure 1: (a) ac transmittivity as a function of temperature at different applied fields for the $Bi_2Sr_2CaCu_2O_8$ sample with CD with a matching field $B_\Phi$=30 G. The temperature-location of the paramagnetic peak, $T_{FOT}$, and the peak height, $\Delta T'$, are indicated. Measurements performed at a ripple field of 0.7 Oe and 3.1 Hz. Inset: Detail of one of the local Hall sensors. (b) Entropy-jump per pancake vortex at the FOT for samples with $B_\Phi$=10 and 30 G (full circles). Open symbols are data from four pristine samples fitted with the model of Ref. [10] for dominant electromagnetic coupling. Dotted lines are fits considering an empirical $T_{FOT}$ dependence (see text).

The data for the four pristine samples are well fitted by the expression $\Delta s \propto 1/(1-(T_{FOT}/T_c)^2)$ (EM coupling fit), shown with a full line in Fig. 1 (b). This functionality was predicted by the theoretical model of Ref.[10] for dominant electromagnetic coupling between pancake vortices. In the case of samples with CD, neither the electromagnetic coupling model, nor the expression for the Josephson dominated coupling, $\Delta s \propto 1/(1-(T_{FOT}/T_c)^2)^{1/2}$, fit the data properly. This disagreement can have its origin in the fact that the model of Ref. [10] neglects the effect of disorder. Indeed, this expression does not fit the data in the case of underdoped $Bi_2Sr_2CaCu_2O_{8+\delta}$ for which electromagnetic coupling is dominant but pinning properties are strongly inhomogeneous. [13] In addition, the authors of Ref. [10] consider the shape of the FOT line proposed by Blatter et al. [12]: in the case of dominant electromagnetic coupling it takes the form $H_{FOT} \sim (1-T_{FOT}/T_c)^2$, but when Josephson coupling dominates, $H_{FOT} \sim (1-T_{FOT}/T_c)^{3/2}$. The shape of the FOT line that we measure in the samples with CD does not agree with either, but is well fitted by an exponent of 1, i.e. $H_{FOT} \sim (1-T_{FOT}/T_c)$. This discrepancy between the shape of the experimentally found FOT line and that proposed by theory can also be at the origin of the EM coupling fit not properly describing the data in samples with CD.

Nevertheless, our data in samples with disorder is well fitted with an empirical expression $\Delta s \propto (1-(T_{FOT}/T_c)^2)^{-\alpha}$. These fits are shown in Fig. 1 (b) with dotted lines. The best fit to our data is obtained with exponents that increase with the density of CD: $\alpha$=1.35 and 1.73 for $B_\Phi$=10 and 30 G, respectively. The origin of this non-unity exponent can be related to the experimental shape of the FOT line differing from that proposed in the model. But the possibility of $\alpha$ being accounted for the effect of the strong disorder introduced by CD can not be ruled out. Further theoretical work is needed in order to explain this empirical analysis.

## 4. Conclusions

The temperature dependence of the entropy jump Δs at the first order transition of the vortex ensemble in irradiated superconducting $Bi_2Sr_2CaCu_2O_8$ samples with a dilute distribution of columnar defects was obtained by measuring the paramagnetic peaks at the transition. Similar to the case of pristine samples, the temperature evolution of Δs diverges on approaching $T_c$. The main effect of a diluted density of CD is a depletion of the entropy jump at lower temperatures. The rate of decrease of Δs enhances with matching field. A model based on EM coupling between pancake vortices does not properly fit the data. An empirical relation between Δs and the FOT temperature is found, but a satisfactory theoretical model taking into account the effects of disorder as well as the measured shape of the transition is still lacking.